\documentclass[twocolumn,pre,a4paper,showpacs]{revtex4}
\usepackage{amsbsy}
\usepackage{amssymb}
\usepackage[dvips]{graphicx}

\begin{document}
Orsay LPT-08-40

\title{Adaptive networks of trading agents}

\author{Z. Burda}
\affiliation{Marian Smoluchowski Institute of Physics
and Mark Kac Complex Systems Research Centre, Jagellonian University,
Reymonta 4, 30-059 Krakow, Poland}

\author{A. Krzywicki}
\affiliation{Univ Paris-Sud, LPT ; CNRS, 
UMR8627, Orsay, F-91405, France.}

\author{O. C. Martin}
\affiliation{Univ Paris-Sud, LPTMS ; CNRS, 
UMR8626, Orsay, F-91405, France.}

\date{\today}

\begin{abstract}
Multi-agent models have been used in many contexts
to study generic collective behavior. Similarly, complex networks
have become very popular because of the diversity of 
growth rules giving rise to scale-free behavior. Here we 
study adaptive networks where the agents trade ``wealth''
when they are linked together while links can appear
and disappear according to the wealth of the corresponding 
agents; thus the agents influence the network dynamics and vice-versa. 
Our framework generalizes a multi-agent model of Bouchand and M\'ezard,
and leads to a steady state with fluctuating
connectivities. The system spontaneously
self-organizes into a critical state where the wealth
distribution has a fat tail and the network is scale-free;
in addition, network heterogeneities lead to enhanced wealth
condensation.
\end{abstract}
\pacs{89.75.Fb, 89.75.Hc, 89.65.Gh}

\maketitle

\section{Introduction}

Multi-agent systems often involve only simple ingredients
and rules, yet can lead
to ``complex'' dynamical behavior. In general the agents of such systems
interact locally (e.g., only within nearest neighbors on a lattice)
or may interact with all other agents (corresponding
to a mean-field system). The network of interactions is then
given a priori and is time independent, only the internal
states of the agents change with time. However there are many
situations where the network structure will be influenced by
the agents' actions: agents may move around and redefine
their neighborhoods, or they may choose their interactions
according to their internal states. For instance, in 
transportation networks, population increases will lead to 
the construction of new links, and inversely
the introduction of new connections will
affect the dynamics of the populations.
Analogous examples abound both in artificial networks (communication,
distribution, etc.) and in natural networks (biological,
ecological, social,...). Having both 
dynamic agents and dynamic connections potentially allows for new phenomena,
be-it at the level of the agents or at the level of their network
of interactions. 

Networks whose links change with time are often referred to as adaptive
networks. There is a rich literature on such networks, reviewed
in particular in refs.~\cite{gross}. But as noted by these authors, 
in most such investigations, the dynamics of the network
occur on a very different time scale from that of the variables
(or ``fields'') affecting these changes. Only for 
consensus-forming networks (see 
for instance \cite{zimmermannEguiluz04,holmeNewman06,kozmaBarrat08})
and variations thereof~\cite{grossLima06}
do the links change at a rate comparable to the fields
(agent opinions in this case); but because opinions are discrete
or because of the nature of these models, one does not 
reach a critical state generically. For our work we seek
systems which spontaneously lead to criticality (without
any parameter fine-tuning) and for which the 
network and the fields driving the network have comparable
time scales.
This challenge is particularly relevant today because 
the last decade has revealed that many
natural and artificial networks have
strong topological heterogeneities and are often
scale free. Surprisingly, the modeling of such networks
is almost always based on growth rules: attachment
of a new link is preferentially done to 
hubs~\cite{barabasiAlbertxx}, or it
depends on fixed hidden variables on the nodes~\cite{parkNewmanxx}.
Such frameworks are out of equilibrium and have no
steady state; in addition, they ignore the dynamics
of the quantities implicitly associated
with the nodes in most real world examples. In the
work described here,
the internal state of each agent can influence
the link dynamics, and at the same time the set of existing links
affects the dynamics of the agents. 

We present our model
in the language of macro-economics where agents have wealth,
perform transactions amongst themselves, and can see 
their wealth increase multiplicatively as in
financial holdings. This choice is motivated by the
overwhelming evidence that wealth 
dynamics in human societies spontaneously evolves to
criticality. In particular, individual wealth follows a 
``Pareto'' law~\cite{pareto89} with power law
tails for the wealthy; similar fat tails also arise
in corporate wealth, e.g., 
in the distribution of sizes of firms~\cite{mandelbrot60,souma01}.

Our framework extends a model proposed by
Bouchaud and M\'ezard~\cite{bouchaudMezardXX}
to the case where the network of interactions
is heterogenous and adaptive. 
We find that in the absence of regulatory mechanisms, 
the system naturally goes to a ``collapsed''
phase where the great majority of agents becomes 
marginalized (poverty stricken) and isolated. 
Including a minimum support level to maintain 
agent connectivity, the system is instead generically driven
to a self-organised critical steady state; in this 
steady state, the
distribution of wealth of agents has a fat tail and the
adaptive network is scale free. 

The paper is organized as follows.
The model is defined in Sect.~\ref{sect_MODEL}; we
also present different observables of interest
and sketch our simulational methods.
In Sect.~\ref{sect_QUENCHED} we exhibit the power laws
arising in the quenched systems: that of the node degree
distribution when the wealth is frozen and that of 
the wealth distribution when the network links are frozen,
considering in particular the effects of network heterogeneity.
In Sect.~\ref{sect_IMPLEMENTATION} the settings of the
respective time scales for agent vs. network dynamics
are presented.
Then we examine the full model where the network is adaptive
(the link dynamics is affected by the agents
and vice-versa) in Sect.~\ref{sect_FULL}.
We conclude in Sect.~\ref{sect_CONCLUSIONS}.

\section{The model}
\label{sect_MODEL}
To specify a multi-agent system, one begins with the possible internal
states of the agents. Since our model builds on that of
Bouchaud and M\'ezard~\cite{bouchaudMezardXX}, 
each of our agents will have its internal state specified 
by a real positive variable, hereafter called its 
``wealth''. Agents see their internal 
state change with time: their wealth will
fluctuate because of returns on investments on the one hand 
and because of exchange of goods against currency on the other; such
exchanges or ``trades'' lead to outflux (from 
purchases) and influx (from sales).
Trades are only performed between linked agents; these links
are either set a priori (``quenched'' or frozen network) or are dynamic
as in adaptive networks. We now explain in detail the dynamics of 
these two parts of our model. 
(Similar ideas have been formulated independently
in ref.\cite{garla}, but, to our knowledge, have not been further developed.)

\subsection{Agent wealth dynamics}

Our system has a fixed number $N$ of agents, whose state at time $t$
is given by $\{ W_i(t) \}_{i=1,\ldots,N}$.
The change in wealth of an agent takes into account trades and returns
on investments. For computational simplicity, we consider a 
discrete time stochastic equation~\cite{bouchaudMezardXX}:
\begin{eqnarray}
\lefteqn{W_i(t+1) = } \nonumber \\ 
& \left( W_i(t) + \sum_j \left(J_{ij}(t) W_j(t) 
- J_{ji}(t)W_i(t)\right)\right) 
e^{\eta_i(t)} \; 
\label{eq:Wdynamics}
\end{eqnarray}
where the parameters $J_{ji}(t)$ describe the fraction of agent
$i$'s wealth which flows to agent $j$ as a result of trading at  
time $t$. The change in an agent's wealth is also affected by the
return on investments
in stock-markets, currency exchange rates, housing or commodity prices etc. 
These investments lead to gains or losses, providing multiplicative changes;
if for example a stock price changes by two percent, 
then the value of a portfolio allocated in that stock
will change by two percent. We model the fluctuations by the
term $\eta_{i}(t)$ which is taken to be a stationary Gaussian variable:
\begin{eqnarray} 
\langle \eta_{i}(t) \rangle & = & 0 \\   
\left\langle \eta_{i}(t)\eta_{j}(t') \right\rangle_c &
= &\sigma_0^2 \; \delta_{ij} \; \delta_{tt'} \ ,
\label{eq:eta} 
\end{eqnarray}
\par
Without the random factors $e^\eta$ in Eq.~(\ref{eq:Wdynamics}),
the total wealth of the system would be conserved; their presence implies
that the total wealth typically grows exponentially with time, 
as discussed in ref.~\cite{bouchaudMezardXX}.
\par
Note that wealth is a relative concept, i.e., independent of
the unit of currency used to measure the $W_i$; hence, 
the wealth dynamics must 
be invariant under the scale transformation 
\begin{equation}
W_i(t) \to \lambda W_i(t) 
\label{scaletransf}
\end{equation}
It is evident that this requirement is satisfied by Eqs.~(\ref{eq:Wdynamics}).
\par
Let us denote by $A_{ij}$ the adjacency matrix of the graph representing
the linking of agents and let us assume, for simplicity, that this graph is
undirected, i.e. $A_{ij}=A_{ji}$. In ref.~\cite{bouchaudMezardXX} Bouchaud 
and M\'ezard have studied in detail the large time behavior in the class of 
models where $J_{ij} \propto A_{ij}$, with a constant proportionality factor
$J_0$, where the graph is time independent. They limited their 
study to fully connected graphs (the model is then analytically solvable) and 
to sparse random (Erd\"os-R\'enyi) graphs. They have shown that in both 
cases the system tends to a steady state where wealth distribution has a power 
law tail at large (relative) wealth values. Furthemore, for sparse random 
graphs and small enough $J_0$ the tail becomes sufficiently fat to lead to 
the ``wealth condensation'' phenomenon: a finite number of agents hold a 
finite fraction of the total wealth, even in the large $N$ limit.
\par
In this paper we propose a two-fold generalization of the study summarized
above. First, we will consider highly inhomogeneous graphs. This is motivated
by the empirical observation that graphs encountered in nature are very often
inhomogeneous. For example, scale-free fat tails of the degre distribution are 
ubiquitous. It is easy to see that for highly inhomogeneous graphs assuming a 
simple proportionality relation $J_{ij} = J_0 A_{ij}$ is untenable. 
Indeed, the loss term in (\ref{eq:Wdynamics}) would then dominate over 
the income term when $W_i$ is large and the rich agents would therefore
prefer to have as few trading partners as possible, contrary to common sense. 
\par
We will assume that all agents trade with the same ``activity'' $J_0$, which is 
constant in time. This means that the total outgoing flow of wealth from the  
agent $i$ equals $J_0 W_i(t)$; in effect, each agent allocates a fixed 
\emph{fraction} $J_0$ of its wealth to trading, a reasonable hypothesis when 
considering life-styles in developped countries.
\par 
For each agent $i$, we shall assume that its outflow of trades (purchases) 
is equally distributed over all agents $j$ it trades with.
Thus, the matrix $J_{ij}(t)$ reduces to
\begin{equation}
J_{ij} = \frac{J_0}{q_j} A_{ij}
\label{JA}
\end{equation}
where $q_i = \sum_j A_{ij}$ is the number of agents trading with $i$. We have
checked, keeping the topology of the graph quenched
(inhomogeneous by construction),
that with Eq.(\ref{JA}) the average wealth is a 
monotonically increasing function of
the node degree: rich agents tend to have many trading partners.
\par
The second generalization we propose concerns the topology of the graph, which
will no longer be assumed frozen. On the contrary, it will adapt itself to the
demands of agents. We now discuss this point in detail.


\subsection{Link dynamics}
\label{subsect_LINKS}

The ``interactions'' between agents are their connections, i.e.,
the support for their mutual 
trades. The corresponding network depends on the internal state of the 
agents themselves, and thus the links between agents are dynamical: they 
can be added or removed over time. To specify these dynamics, we shall 
model the time evolution of the adjacency matrix $A_{ij}(t)$, which is
now assumed to be time dependent:  $A_{ij}(t)=1$ if at time $t$
the agents $i$ and $j$ can trade with each other and $A_{ij}(t)=0$ otherwise. 
\par
We have to define the dynamics for the graph evolution
$A_{ij}(t) \rightarrow A_{ij}(t+1)$. To model its dependence on 
wealth distribution, we propose a preferential trading rule, 
according to which the probability of establishing a new trade 
connection between two agents is roughly proportional to the wealth
of each agent. To turn this rule into a probabilistic recipe 
one has to define a quantity in the range $[0,1]$ which can be 
interpreted as a probability. Instead of $W_i(t)$,  we will use
normalized quantities which express the wealth of agents 
in units of the current mean value of the wealth in the ensemble:
\begin{equation}
w_i(t)=\frac{W_i(t)}{\overline{W}(t)} \ , \qquad 
\overline{W}(t) = \frac{1}{N}\sum_i^N W_i(t)
\label{wW}
\end{equation}
Clearly $w_i(t)$ is invariant under the scale transformation 
Eq.~(\ref{scaletransf}).
The position, or solvency, of the agent in the system is better
reflected by its normalized wealth than by its absolute
wealth. In these units the mean value of wealth 
is by construction always equal to unity, $\overline{w} = 1$. In our wealth
preferential trading rule, the probability of establishing a new 
trading connection, $A_{ij}(t)\!=\!0\rightarrow A_{ij}(t+1)\!=\!1$,
increases with $a w_i(t) w_j(t)$ where $a$ is some
proportionality factor. The only problem is that even if $a$ is small, 
this quantity may exceed one for large $w_i$ and $w_j$ and thus loose
a probabilistic interpretation. To avoid this pathology we set:
\begin{equation}
\mbox{\rm Prob}(\mbox{add link} \ ij) = 
\frac{a w_i(t) w_j(t)}{1+a w_i(t)w_j(t)}
\label{add}
\end{equation}
Of course trade connections between agents do not necessarily 
exist for ever. We allow in 
our model for the possibility of abandoning an existing trade connection,
$A_{ij}(t)\!=\!1\rightarrow A_{ij}(t+1)\!=\!0$.
For simplicity we shall assume that the probability of breaking the trade 
or equivalently of removing an existing link between $i$ and $j$ is constant 
in time and independent of the agents' wealth:
\begin{equation}
\mbox{\rm Prob}(\mbox{remove link} \ ij) = r \ll 1
\label{rem}
\end{equation}
Taken together, Eqs.~(\ref{add})-(\ref{rem}) along with 
Eqs.~(\ref{eq:Wdynamics}) define an adaptive 
network, preserving the property of 
invariance under Eq.~(\ref{scaletransf}) of the original Bouchaud-M\'ezard model.
\par
The model is now formulated. As will be seen, it displays a very interesting 
pattern of adaptation of the network topology to the wealth distribution and 
vice versa. Before we discuss these properties, let 
us first consider the limiting cases 
in which only one sector is active while the other is quenched:
(a) the network topology evolves according to the dynamics described 
above while the wealth distribution is quenched; 
(b) the wealth distribution evolves according to the dynamics described 
above while the network topology is quenched.

\section{Quenched dynamics}
\label{sect_QUENCHED}

\subsection{Quenched wealth distribution}
\label{subsect_QUENCHED_WEALTH}


Assume now that the distribution of wealth is constant
during the evolution of the network. The process
of adding and removing links between nodes $i$ and $j$ 
can be viewed as a two-state Markov chain. Since the
weights are constant in time $w_i(t)=w_i$, the probability 
of adding the link $ij$ (cf. Eq.~\ref{add}) is constant as well.
Similarly, the probability of removing the link $ij$ is constant 
(cf. Eq.~\ref{rem}). One can then easily determine the stationary 
probability for this Markov chain; one finds that
for this stationary distribution
the probability that there is a link between nodes $i$ and $j$ equals
\begin{eqnarray}
p_{ij}& = &\frac{\mbox{\rm Prob}(\mbox{add link} \ ij)} 
{\mbox{\rm Prob}(\mbox{add link} \ ij) + 
\mbox{\rm Prob}(\mbox{remove link} \ ij)} \nonumber\\
& = &\frac{\beta w_iw_j}{1+\beta(1+r)w_iw_j}  \; ,
\label{pij}
\end{eqnarray}
where $\beta=a/r$. Assume that 
the weights $w_i$ are independent identically distributed random numbers
with some probability distribution $\rho(w) dw$ such that
the mean is 1, i.e., $\langle w \rangle = \int w \rho(w) dw = 1$. In
this case one can easily see that the total
expected number of links of the network can be
bounded from above:
\begin{equation}
\langle L \rangle = 
\frac{N(N-1)}{2} \langle p_{ij} \rangle \le 
\beta \frac{N(N-1)}{2} 
\label{upl}
\end{equation}
We used the fact that the denominator of $p_{ij}$ is
by construction equal or larger than one and 
$\langle w_i w_j \rangle \approx \langle w_i \rangle \langle w_j \rangle = 1$.
Additionally if the coefficient $\beta$ is inversely 
proportional to the number of nodes, i.e., $\beta= Q/N$,  
the network will be sparse and
the expected number of links will
approach the upper bound given in (\ref{upl})
in the limit $N \rightarrow \infty$ because  
the denominator will tend to one. Thus, the mean 
connectivity of the network is expected to be
\begin{equation}
\overline{q}  = \frac{2 \langle L \rangle}{N} \rightarrow Q
\label{qQ}
\end{equation}
for $\beta = Q/N$ and $N \rightarrow \infty$. For
$r \ll 1$ the probability (\ref{pij}) that there is a link 
between a pair of vertices $i$ and $j$ is for all practical purposes the same as 
in the Park-Newman model~\cite{parkNewmanxx}, so we expect that the two
models will 
behave similarly for small $r$, and in fact we have checked
that this is indeed the case. 
\par
It is known from the considerations of Park and Newman \cite{parkNewmanxx} 
that if $w_i$ are independent identically distributed random numbers 
with a probability distribution having for large $w$ a scale-free tail 
$\rho(w) dw \sim w^{-1-\mu} dw$ with $\mu>1$ then the node degree 
distribution also exhibits the scale-free behaviour 
$\mbox{\rm Prob}(q) \sim q^{-1-\mu}$ (in a range of values of $q$) provided 
the network is sparse. This is what we observe too.
\par
The original Park-Newman model used the concept of fitness, 
closer in spirit to the unnormalized weights $W_i$ rather
than the normalized ones $w_i$ (\ref{wW}). The main difference 
between the two frameworks is that the average fitness $\overline{W}$ 
for the ensemble of $N$ numbers $W_i$, $i=1,\ldots,N$ 
may differ from ensemble to ensemble while for the normalized weights
by construction it is always constant $\overline{w}=1$. In effect,
if one substitutes $w$'s by $W$'s and $\beta \rightarrow \beta_{PN}$
in (\ref{pij}) and neglects $r$ to get the original Park-Newman 
model, one obtains a simple relation between the two definitions of $\beta$:
\begin{equation}
\beta = \beta_{PN} \overline{W}^2 
\label{eff_beta}
\end{equation}
Note that in the Park-Newman model, 
$\beta_{PN}$ is constant; then the above identification
leads to a $\beta$ that
fluctuates from event to event as a result of the fluctuations of 
the average $\overline{W}$. 
\par
For large $N$, by virtue of the central limit theorem, $\overline{W}$ is, 
for $\mu > 2$, a Gaussian random number fluctuating around the mean
$\langle W \rangle$ within a range of size $\sim N^{-1/2}$. 
For $1<\mu<2$, $\overline{W}$ is a L\'evy random number whose 
probable deviations from the mean are of order $\sim N^{1/\mu-1}$, 
Finally, for $\mu<1$, $\overline{W}$ is a L\'evy random number of 
order $N^{1/\mu-1}$, subject to enormous fluctuations.
In other words, as long as $\mu>1$, the Park-Newman construction and ours
differ for large $N$ by a trivial rescaling (\ref{eff_beta}),
while for $\mu<1$ the mapping breaks down.
\par
Our network evolution has been defined using ``computer'' time. Hence, if
$\epsilon$ denotes the unit of the physical time, the parameters
$a$ and $r$ are both proportional to $\epsilon$. However, as
was shown above, as long as $r \ll 1$ the relevant control parameter 
of the model, as far as the topology of the network is concerned, is the 
ratio $\beta=a/r$, which is insensitive to the value of $\epsilon$.
However, the value of $r$ controls the rate of updates of the algorithm
and, therefore, the autocorrelations during the history of a computer
simulation. We set $r=0.1$ in our numerical work, considering $a$ 
as the relevant adjustable parameter.

\subsection{Quenched network}
\label{subsect_QUENCHED_NETWORK}

\subsubsection{The continuous time limit}
Now assume that the network is fixed during the evolution of 
weights: $A_{ij}(t)=A_{ij}$. In this case (cf.
ref.~\cite{bouchaudMezardXX}) 
Eq.~(\ref{eq:Wdynamics}) has a continuous time limit under a proper
scaling of the parameters of the model. Let $\tau=\epsilon t$ denote
the physical time and set
\begin{eqnarray}
J_0 = \epsilon J \\
\sigma_0 = \sqrt{\epsilon} \sigma
\end{eqnarray}
In the limit $\epsilon \rightarrow 0$ one gets from (\ref{eq:Wdynamics})
together with (\ref{JA}) the following stochastic equations 
(in the Stratonovich sense):
\begin{eqnarray}
\label{continuum}
\lefteqn{\frac{dW_i(\tau)}{d\tau} = } \\
& \sigma \frac{dB_i(\tau)}{d\tau} W_i(\tau) + 
J \sum_{ij} \left(A_{ij} W_j(\tau)/q_j - A_{ji} W_i(\tau)/q_i\right) \nonumber
\end{eqnarray}
where $B_i(\tau)$ is a $N$-dimensional Wiener process. Dividing both sides
by $\sigma^2$ and rescaling the time variable $\tau \rightarrow \sigma^2 \tau$
one sees that at large time the only relevant parameter 
is $J/\sigma^2$.
\par
We simulate the model on a computer using its discrete formulation. However,
we try to be close to the continuous time limit, setting $\epsilon$ very small
(in our runs we used $\epsilon = 0.001$). Since with such a choice 
one expects that the dynamics depends
on $J/\sigma^2$ only, we can without any loss of generality set the physical
parameter $\sigma=1$. 
\par
When the graph is complete, 
that is for $A_{ij} = 1 - \delta_{ij}$,
Eq.~(\ref{continuum}) can be solved analytically \cite{bouchaudMezardXX}.
For $J > 0$, $N \rightarrow \infty$ and $\tau \rightarrow \infty$
one gets a stationary distribution for the normalized weights (\ref{wW}).
It has a fat tail $\sim w^{-\mu-1}$ at large $w$, with the exponent
$\mu = 1 + J/\sigma^2$. Notice, that for $J=0$ the
stationary solution does not exist, and therefore the limit 
$J \rightarrow 0$ is singular.
The authors of ref.~\cite{bouchaudMezardXX} have 
also shown, using numerical simulations, 
that for sparse random Erd\"os-R\'enyi graphs, one again gets a fat tail but 
with an exponent $\mu$ smaller than one if $J/\sigma^2$ 
is smaller than a certain critical value. We have repeated these
simulations 
in our version of the model for a sample 
of network topologies. We observe that 
the fat tail always emerges and that the 
corresponding exponent depends weakly on 
network topology (see later). The occurence of such a fat tail 
with $\mu<1$ in the wealth distribution has consequences that we now 
discuss in detail.

\subsubsection{Poverty and wealth condensation}
\label{subsect_POVERTY}
Let us carefully study the consequences of using the normalized
$w$'s instead of $W$'s (our discussion is inspired by ref. 
\cite{bouchaudMezardXX2}). For the sake of simplicity, but without any real
loss of generality, assume that the probability distribution
of $W$ is (we omit the index $i$ for simplicity of writing):
\begin{equation}
\mbox{\rm Prob}(W) dW = \mu W^{-\mu -1 } dW \ , \quad \ W \ge 1
\label{rho0}
\end{equation}
and zero otherwise. Assume first that $\mu>1$ so that the mean
$\langle W \rangle$ is well defined. We want 
to calculate the probability distribution 
of the scaled variable defined in (\ref{wW}):
\begin{equation}
w = \frac{N W}{W+S}
\label{S}
\end{equation}
where $w=w_i$, $W=W_i$ and $S=\sum_{j\ne i} W_j$ is
the sum of remaining terms. For large $N$ one can replace
$S$ by its mean value $S = sN$, where $s = \langle W \rangle = \mu/(\mu-1)$. 
After trivial algebra one gets from (\ref{rho0})
\begin{eqnarray}
\lefteqn{\mbox{\rm Prob}(w) dw = } \nonumber \\
& C w^{-\mu-1} 
\left(1 - \frac{w}{N}\right)^{\mu-1} dw \ , \quad  
w \in [w_{min}, w_{max}]
\label{rho}
\end{eqnarray}
\noindent
where $w_{max} = N$, $w_{min} = s^{-1}$ and $C= \mu s^{-\mu}$.
\par
The above distribution has natural cut-offs, as expected. 
In addition to the behavior $w^{-\mu-1} dw$ inherited
from (\ref{rho0}) it involves a factor $\left(1 - \frac{w}{N}\right)^{\mu-1}$
suppressing $w$'s of order $N$. The lower 
cut-off $w_{min} = s^{-1} = (\mu-1)/\mu$ 
is finite as long as $\mu>1$. For $\mu \leq 1$, one has to redo the analysis. 
\par
Let us observe that strictly speaking $s$ 
is not fixed but fluctuates. Hovever,
when $\mu>1$ its departures from the average can be neglected when $N$ is
large enough. 
When $\mu < 1$ this is no longer true. If $\mu \le 1$ the sum $S$ (\ref{S}) 
does not increase linearly with $N$: instead $S$ scales as $\eta N^{1/\mu}$, 
where $\eta$ is some constant, which shall be calculated below. So in 
this case the lower cut-off 
$w_{min}$ in (\ref{rho}) is
\begin{equation}
w_{min}= \eta^{-1} N^{1-1/\mu}
\end{equation}
as one can see by inserting $W_{min}=1$ on the right 
hand side of (\ref{S}). The cut-off goes
to zero as $N\rightarrow \infty$, but for any finite $N$ it is finite. 
It is essential to keep it finite while calculating the integral 
$\int \mbox{\rm Prob}(w) dw$ since otherwise 
the singularity $w^{-1-\mu}$ at zero 
would make the integral (\ref{rho}) diverge. With 
$C=\mu \eta^{-\mu} N^{\mu-1}$ (for $\mu<1$) the integral
is properly normalized $\int_{w_{min}}^N \mbox{\rm Prob}(w) dw =1$ for 
$N\rightarrow \infty$ and the mean value of $w$ is 
$\langle w \rangle = \int_0^N w \rho(w) dw =
\eta^{-\mu} \mu \Gamma(\mu) \Gamma(1-\mu)$.
(In the calculation of the mean value $\langle w \rangle =1$ one can set
$w_{min}=0$ since the singularity at zero is integrable). 
Hence, $\langle w \rangle = 1$ if $\eta^\mu = \mu \Gamma(\mu) \Gamma(1-\mu)$.
\par 
One can calculate the probability that $w$ is smaller than
a given small fixed number $\Delta w$:
\begin{equation}
\mbox{\rm Prob}(w < \Delta w) = 
\int_{w_{min}}^{\Delta w} \rho(w) dw \approx 1 - 
c N^{\mu-1}
\end{equation}
where $c = (\sigma \Delta w)^{-\mu}$, so that $\mbox{\rm Prob}(w < \Delta w) \rightarrow 1$ 
for $N\rightarrow \infty$. This means if one makes a fixed-bin
histogram of $w_i$'s for large $N$, then almost all $w_i$'s
will be in the first bin adjacent to zero. This phenomenon can be
called a ``poverty condensation''.
\par
Another surprising feature 
of the wealth distribution when $\mu<1$ is that the factor
$\left(1 - \frac{w}{N}\right)^{\mu-1}$ does not introduce
a suppression of $w$ of order $N$, but an enhancement.
The singularity at $w=N$ is integrable. Intuitively this means 
that in a large sample of $w_i$'s, most values are concentrated at 
zero, but a few remaining ones are of order $N$. This is also what
one can infer from the calculation of the inverse participation ratio 
$Y_2$~\cite{bouchaudMezardXX,bouchaudMezardXX2}. For $N\rightarrow \infty$ 
\begin{equation}
\langle Y_2 \rangle = 
\sum_i^N \left( \frac{w_i}{N}\right)^2 = \frac{1}{N} \langle w^2 \rangle =
1-\mu.
\label{invparticip}
\end{equation}
is a finite positive number when $\mu<1$ whereas $Y_2=0$ for $\mu>1$. This 
shows that in a large sample of $w_i$'s a finite fraction of 
them is of order $N$. This is the 
``wealth condensation'' signaled by Bouchaud 
and M\'ezard. Notice, that poverty and
wealth condensation occur simultaneously.
\par
\begin{figure}
\includegraphics[width=8cm]{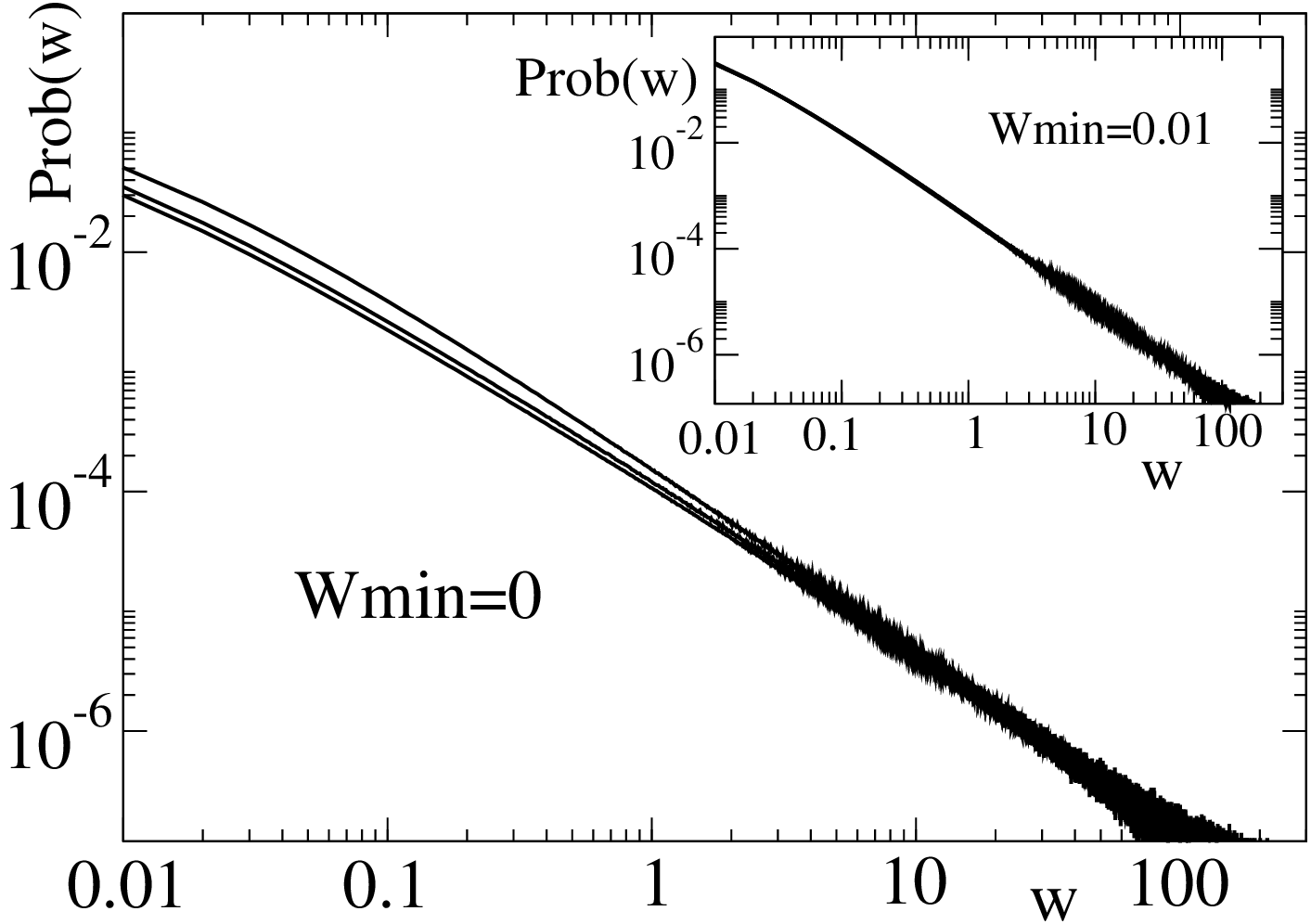}
\caption{\label{fig:wealthFatTailsQuenched} 
The distribution of wealth has a power law tail
in generic networks; furthermore the corresponding
exponent is not sensitive to the network structure
as shown here for different kinds of networks (Erd\"os-R\'enyi,
scale-free or with fixed connectivity). The network size 
is $N=1000$, the coupling $J=0.005$ and the fitted 
slopes equal $1+\mu=1.447(2)$ to $1.465(2)$. Inset: the same, but
after imposing a lower cut-off $W_i>0.01$ on agent's wealth. 
Here, the essentially common slope is $1+\mu=1.691(1)$.}
\end{figure}

The above discussion refers to simple sampling of $w_i$'s. Now let the
agent's
wealth be dynamic (but still keeping the geometry frozen).
We show in Fig.~\ref{fig:wealthFatTailsQuenched} the wealth distribution 
calculated keeping the network
quenched, for Erd\"os-R\'enyi, scale-free with exponent $1.5$ and regular
networks with fixed connectivity (in all these cases we set the average
connectivity to 4). The parameter $J$ is set to $0.005$. The fitted 
slopes equal $1+\mu=1.447(2)$ to $1.465(2)$. In 
agreement with the above discussion,
most of the agents (about 80\%) are concentrated in the leftmost bin $[0,0.01]$.
This completes the discussion with either the wealth or the links 
frozen. From now on we focus on the full model.
%
\section{Agent and network time scales}
\label{sect_IMPLEMENTATION}
For our simulations, we alternate the updatings of the wealth and links. In
one update of the wealth, 
Eq.~(\ref{eq:Wdynamics}) is used for each node.
Once all new $W_i$ are found, they are renormalized, so that $\sum_i W_i=N$.
In one update of the geometry we pick a pair of nodes at random and 
use Eqs.~(\ref{add}) or (\ref{rem}), when 
the nodes are connected or not, respectively.
This is repeated $N(N-1)/2$ times.
But this poses the problem of the relative frequency of the updates, i.e.,
what are the two associated time scales for wealth and
link updates. In physical systems, these time scales are a priori
given by the laws of physics. One example of this is
the coupling of matter and geometry in theories of gravity. Network nodes
involve ``matter'' fields while the network links describe the curved
geometry of interactions. The theory involves coupling constants
which specify the dynamical time scales of matter and geometry
degrees of freedom. Comparing to our agent based model, matter is analogous
to wealth and geometry is described by the network topology.
\par
For our adaptive network model of agents, how should one set
the two time scales? As pointed out
in a recent review \cite{gross}, in most models studied so far 
the wealth changes
either much faster or much more slowly than 
the geometry. We wish to have the two
time scales be comparable. Once the value of $\epsilon$ has been chosen,
the rate of wealth updates is fixed. As already 
mentioned, the rate of geometry updates
is controlled by the parameter $a$ or equivalently $r$. To compare the two 
rates, we have randomized the system and then let it evolve keeping the 
wealth or the geometry quenched. We found that (with our choice 
of $\epsilon$ and $r$) the autocorrelation length 
for wealth is two orders of magnitude larger than for
geometry. Consequently, in the simulations of the coupled system 
we alternate 1 sweep of the geometry with 100 sweeps of the wealth
(and there are about ten updates of the whole system within one 
autocorelation time interval).
\par
The physical control parameters are $J$ and $\beta$. Actually, as will be 
seen, the choice of $\beta$ has little influence on the wealth distribution;
it controls the average degree of the network. The degree distribution itself
turns out to have a smooth dependence on $\beta$ when it is plotted versus the scaled
variable $q/\langle q\rangle$. On the other hand the value taken by $J$ 
is essential for the behavior of the system.
\par
The ansatz Eq.~(\ref{JA}) generates a positive correlation between the degree
of a node and the wealth stored in this node. One can suspect that this
leads to a breakdown of ergodicity for heterogeneous networks. And indeed, 
ergodicity is broken as
long as the geometry is quenched: if at a certain moment a given agent
is the poorest (richest) it never becomes the richest (poorest) during
the run history. We have found, however, that the ergodicity is restored
when wealth and links get coupled. In a sense, this coupling increases the 
``social mobility''.

\section{Adaptive network of interacting agents}
\label{sect_FULL}

\subsection{Network collapse in the absence of a cut-off}
\label{subsect_CUTOFF}
The poverty condensation has dramatic consequences when one couples wealth
to geometry. As soon as one enters the regime where the wealth distribution 
develops a fat tail with $\mu < 1$, nearly all nodes become progressively 
isolated (have zero degree) and all wealth 
becomes the property of a tiny minority. A 
modification of the rules is called for, either for wealth (welfare) or for 
connectivity (not considered here). We impose a lower cut-off on $W_i$'s viz. 
$W_i > W_{min}=0.01$. Since we work with scaled variables $w_i$ and since we 
recalculate them after each wealth update, the 
$w_i$'s inherit a similar cut-off, 
except that it somewhat smeared around $0.01$. 
In the inset of Fig.~\ref{fig:wealthFatTailsQuenched} 
we show the wealth distribution for quenched 
networks when this cut-off is imposed; no collapse is possible there.
Hence, a calculation with and without 
cut-off can be compared and one notices 
that the fat tail appears in both cases, 
although the exponent $\mu$ is a little larger when the cut-off is present.
When the network is adaptive, the cut-off prevents the collapse.

\subsection{General overview}
\label{subsect_GENERAL}
Before presenting more datailed data on the wealth and degree distributions
and on the correlation between the two, let us have a general view of
the model's properties. 
\par
With an ongoing trading activity and link changes, the system evolves and
empirically always seems to reach a steady state that is unique (independent
of the initial conditions). Furthermore, there is a smooth large
volume limit.
\par
\begin{figure}
\includegraphics[width=8cm]{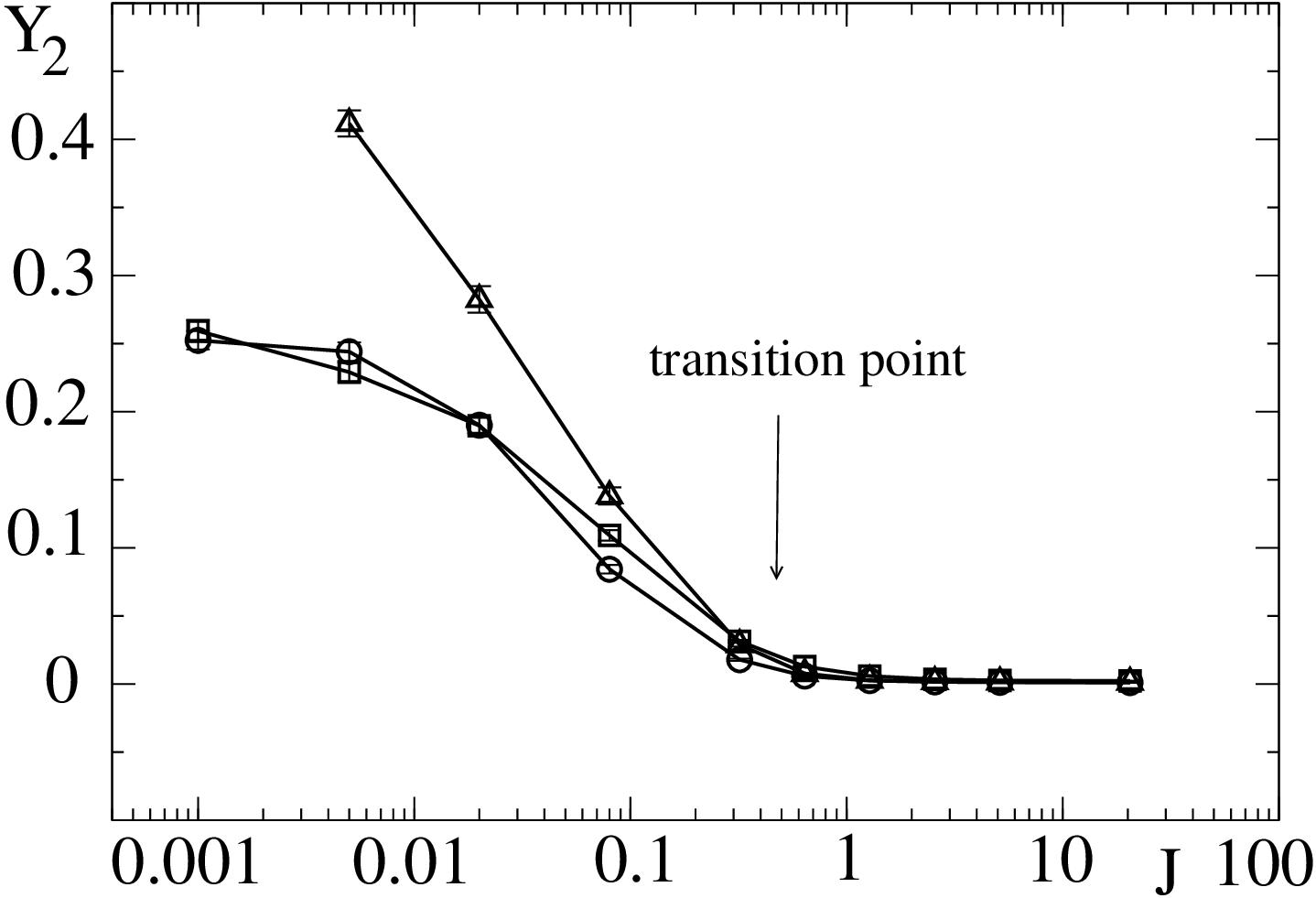}
\caption{\label{fig:Y2w} 
The inverse participation ratio is an order parameter
for wealth condensation. One goes from a homogeneous phase
at large $J$ to a condensed phase at low $J$ where a finite
number of agents hold a finite fraction of the total
wealth (we have set $\sigma=1$). 
Shown are data for adaptive networks with $\beta=0.020$ (squares)
and $\beta=0.20$ (circles) when $N=1000$.
The analogous data for quenched random networks
with $\langle q\rangle=4$ are displayed using triangles.
The lines are here to guide the eye. Note that the transition point 
is insensitive to the type of network: quenched or adaptive.}
\end{figure}

It is most instructive to examine the dependence of the
inverse participation ratio $Y_2$ defined in 
Eq.~(\ref{invparticip}) versus $J/\sigma^2$
(cf. Fig.~\ref{fig:Y2w}). The qualitative 
behavior is similar to that observed in the
Bouchaud-M\'ezard model. (For completeness we 
show also in the figure the data corresponding to 
a calculation with quenched random network.) We find that $Y_2$ 
is finite as long as $J/\sigma^2$ is small enough, 
it falls progressively as $J/\sigma^2$ increases and 
eventually settles at a value of order $1/N$ when 
$J/\sigma^2$ is increases beyond a certain critical 
value. Notice that an increase of $\beta$ 
from $0.020$ to $0.20$ has very little effect. Remember 
also that $Y_2=1 -\mu$ 
as long as the distribution has a tail falling off as a power
with $\mu<1$ (evidence for 
this scale-free behavior will be presented 
in Sect.~\ref{subsect_WEALTH}). Hence, 
the evolution of the wealth distribution slope with 
$J/\sigma^2$ can be immediately deduced from Fig.~\ref{fig:Y2w}. 
\par
The model has two distinct phases. An educated guess is that in the large 
$J/\sigma^2$ phase the dynamics is qualitatively well described by the 
``mean field'' approximation of ref. \cite{bouchaudMezardXX}. This is also 
suggested by the simulations we have carried out, which are however strongly 
affected by finite-size corrections (the efficiency of our algorithm does not 
allow us to go far beyond $N=1000$). The low  $J/\sigma^2$ phase 
is by far more interesting and we focus on it hereafter. In the following paragraphs, we shall 
consider successively network properties, wealth properties and joint effects.
%
%
\subsection{Scale-free steady-state networks}
\label{subsect_NETWORKS}
\begin{figure}
\includegraphics[width=8cm]{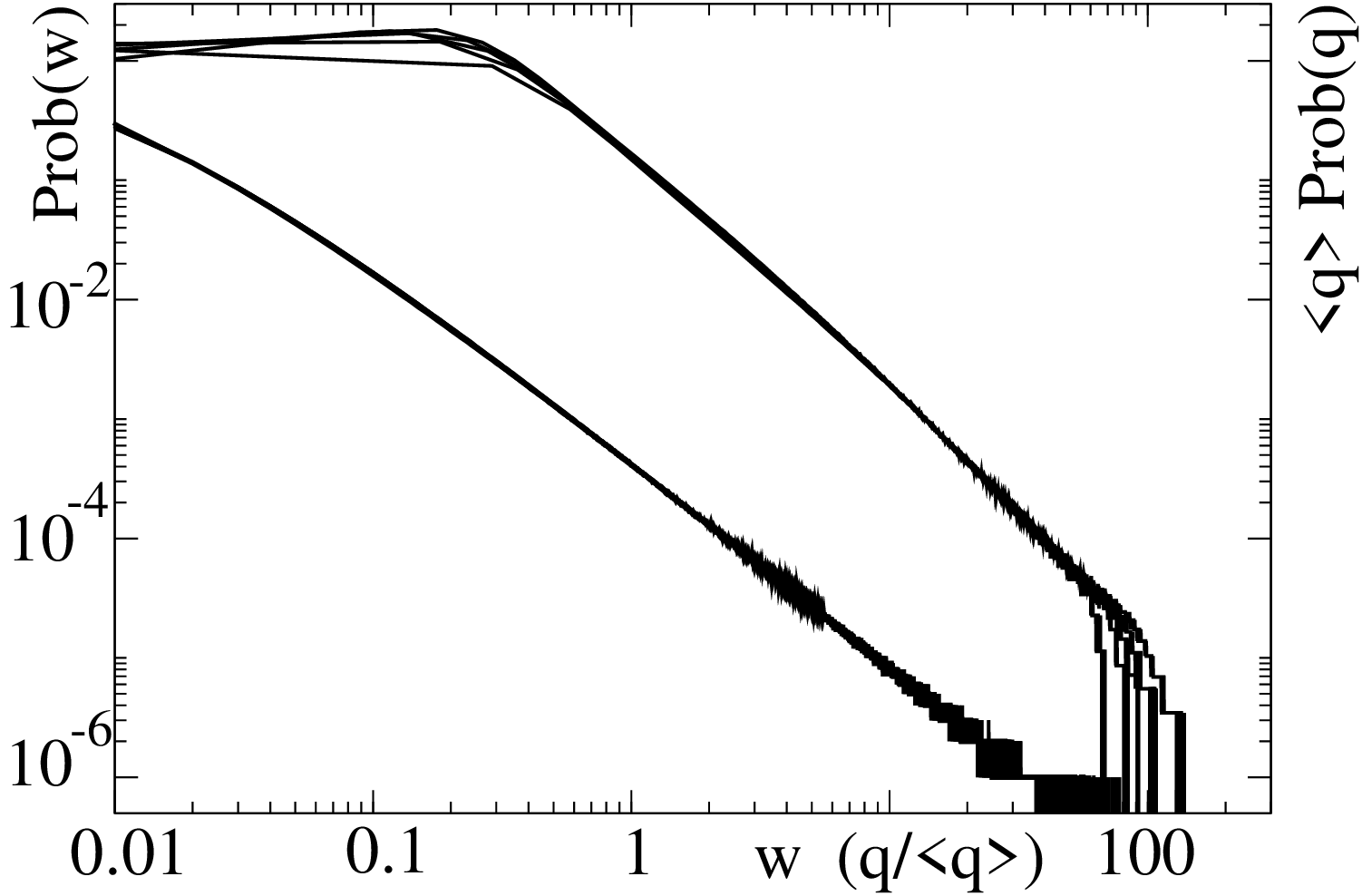}
\caption{\label{fig:betadependence} 
Adaptive networks: wealth (left) and degree (right)
distributions for $N=1000$, $J=0.005$ and $\beta$ ranging from
$0.020$ to $0.120$. The slopes are $1+\mu=1.644(2)$ and $\gamma=2.105(5)$ 
respectively. The lines are to guide the eye. 
The scale-free shape of both distributions is
evident. We have plotted the degree distribution using the rescaling 
$q \to q/\langle q \rangle$.}
\end{figure}

\begin{figure}
\includegraphics[width=8cm]{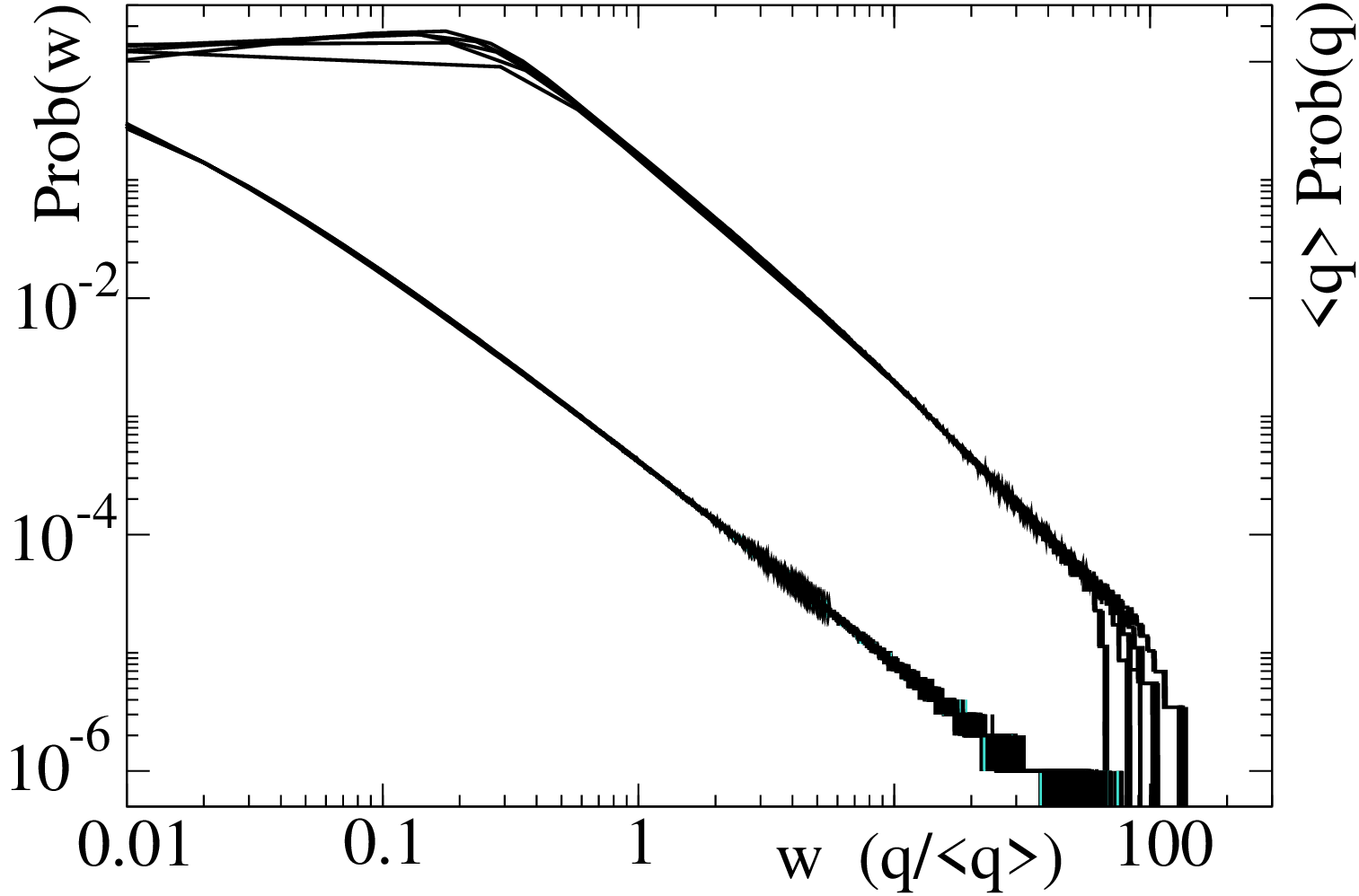}[b]
\caption{\label{fig:Jdependence} 
Adaptive networks: wealth (left) and degree (right)
distributions for $N=1000$,  $\beta=0.020$ and $J$ ranging in small
steps from $0.001$ to $0.010$. The lines are to guide the eye.}
\end{figure}

We display in Figs.~\ref{fig:betadependence}-\ref{fig:Ndependence} 
the distribution of node connectivities
$q$ in the case of sparse networks (cf. Eq.~(\ref{qQ})). 
%
For not too large $J$, the degree distribution depends weakly
on the value of this parameter whereas the dependence on $\beta$
is rather strong. However, scaling the degree 
$q \to q/\langle q \rangle$ we find 
that the tail of the degree distribution is both scale free and 
insensitive to $N$ at large $N$. Such scale-free behavior
seems to be generic; indeed we find it for all the parameter values
we have explored.
%
Thus, the tail of the distribution of $q$ behaves as
\begin{equation}
\mbox{\rm Prob}(q) \sim q^{-\gamma}
\end{equation}
where $\gamma$ depends on the values of 
the control parameters though it is not sensitive to them.
Furthermore, we find that $\gamma$ does not
go below $2$ so no node carries a finite fraction of
all links. This can be referred to as lack of
``link condensation''. One can define $Y_2$ for the
degree distribution by replacing $w_i \to Q_i=Nq_i/2L$ in
the defining equality in (\ref{invparticip}) (notice that $\sum_i Q_i =N$).
One finds that this $Y_2$ is typically one order of magnitude smaller
than the corresponding parameter for the wealth.

\subsection{Power-law wealth distributions}
\label{subsect_WEALTH}

\begin{figure}
\includegraphics[width=8cm]{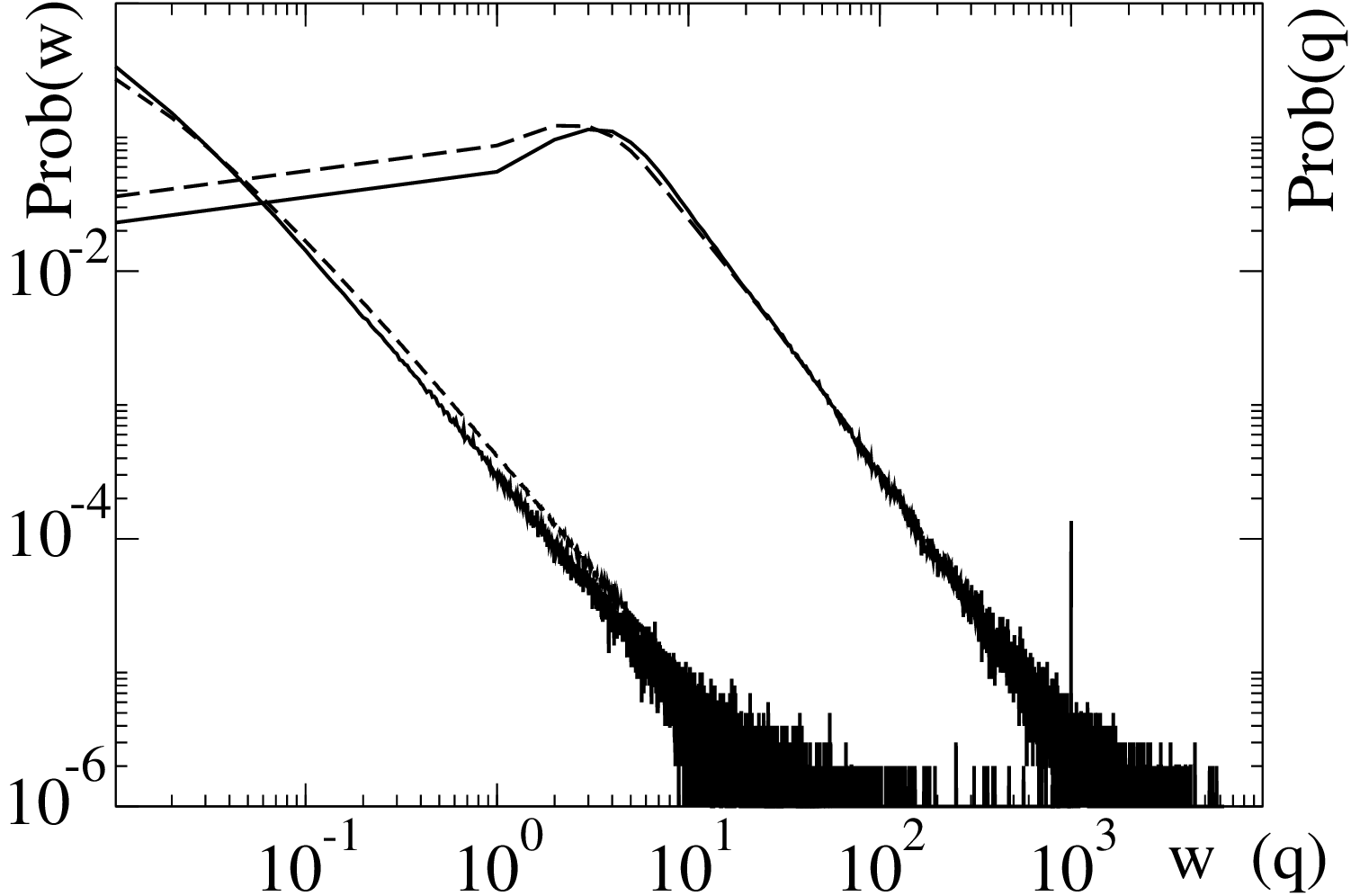}
\caption{\label{fig:Ndependence} 
Adaptive networks: wealth (left) and degree (right)
distributions for $J=0.005$ at $N=1000$ (dashed line; $\beta=0.20$) and 
$10000$ (solid line; $\beta=0.02$). For wealth  
the slopes are $1+\mu=1.697(1)$ and 
$1+\mu=1.749(9)$ at $N=1000$ and $10000$, respectively. The slope 
for the tail of the degree distribution is $\gamma=2.069(13)$. 
The figure illustrates that the exponents depend very weakly on 
the network size as expected in a thermodynamic limit.}
\end{figure}

Now we focus on the properties of the agents' wealth. We saw
that when the network was quenched, a fat tail appeared
generically so it will come as no surprise that in the
adaptive network model the distribution of wealth $\mbox{\rm Prob}(w)$
again has power law tails. Examples of such tails are given 
in Figs.~\ref{fig:betadependence}-\ref{fig:Ndependence} for the case of 
sparse networks ($\beta$ scaling as $1/N$).
As already mentioned, the exponent $\mu$ depends
on the parameters of the model, weakly on $\beta$, more strongly 
on $J$, as can be deduced from the curves in Fig.~\ref{fig:Y2w}.

\subsection{Wealth and topology are associated}
\label{subsect_WEALTH2}
\begin{figure}
\includegraphics[width=8cm]{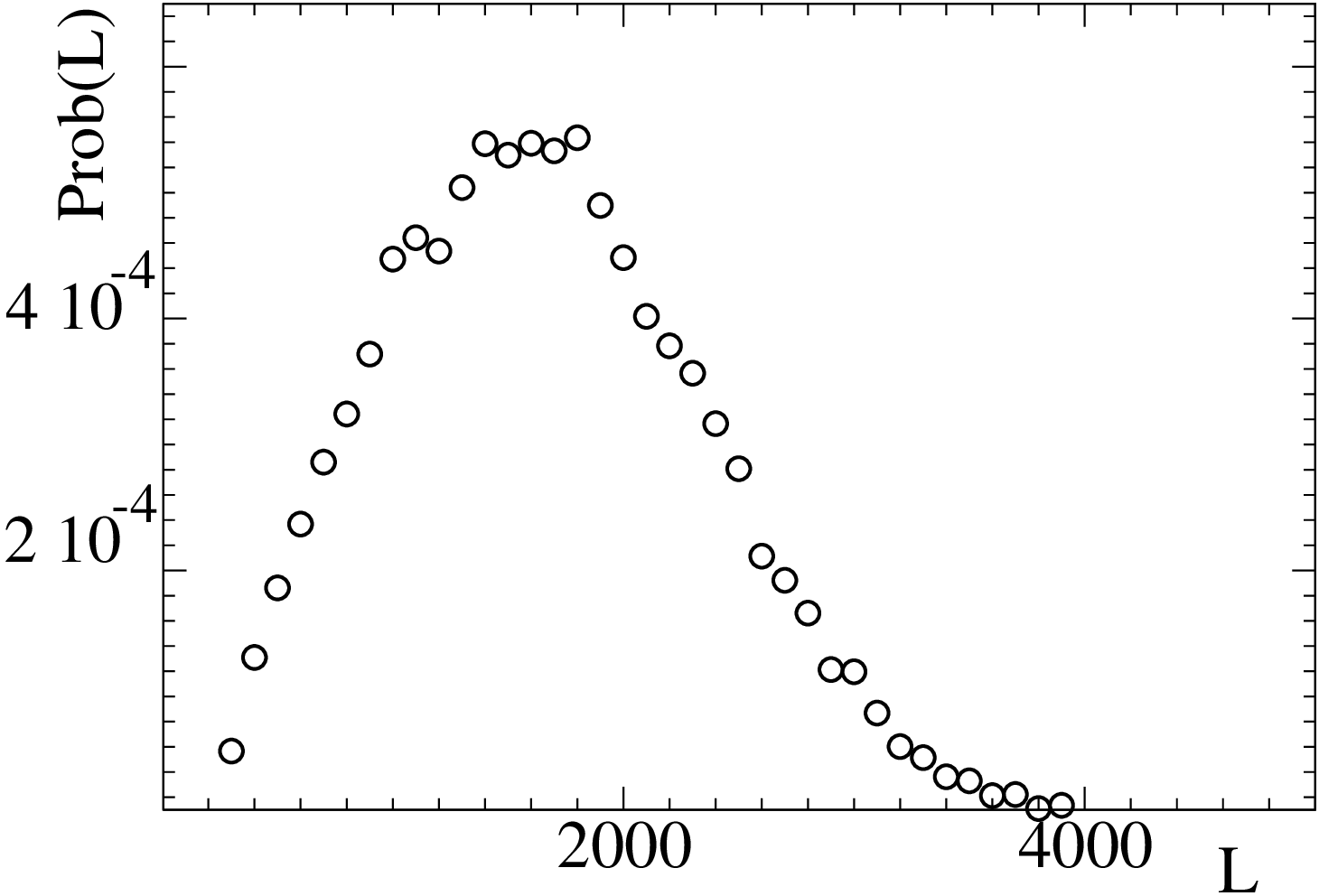}[b]
\caption{\label{fig:linktotalNumber} 
The number $L$ of links in the network
fluctuates substantially in the steady state. Here 
$N=1000, J=0.005$ and $\beta=0.020$, with a binning of size 100.}
\end{figure}
The relative insensitivity of our results to 
parameter changes might suggest that the steady
state reached at large time by the system is extremely stable. It turns out,
however, that the system is actually subject to very large fluctuations,
for instance for the total wealth, and that these fluctuations are much
larger than those observed when the geometry 
is kept quenched. This can be traced
back to the slow fall-off of the wealth distribution: with such $w_i$'s the
link dynamics of Sect.~\ref{subsect_LINKS} necessarily generates networks with
strongly fluctuating number of links. An illustrative 
example is given in Fig.~\ref{fig:linktotalNumber}, which 
shows that the total number of links has a fairly broad distribution.
%
\begin{figure}
\includegraphics[width=8cm]{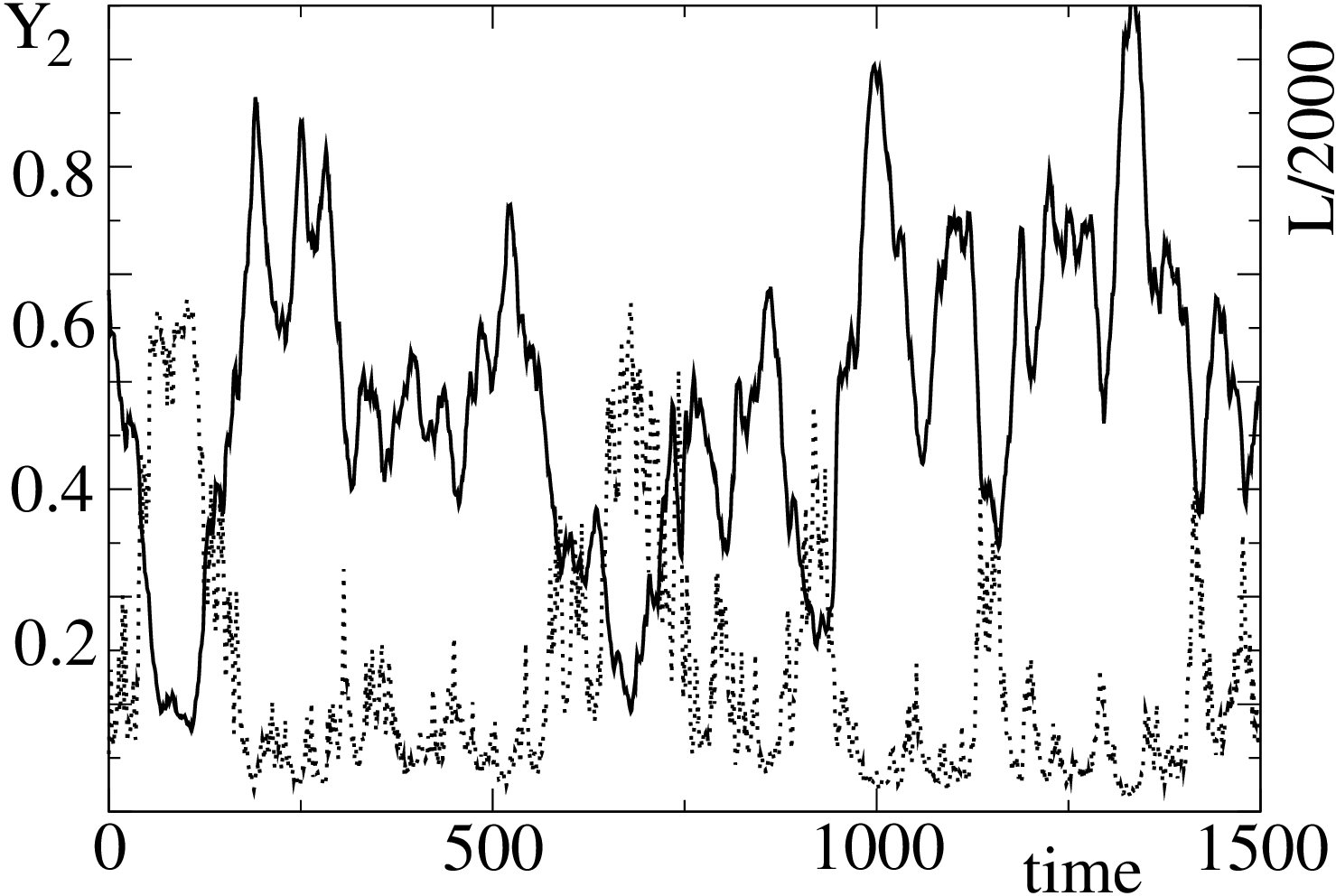}
\caption{\label{fig:Y2vsL} 
Adaptive networks: wealth inverse participation ratio 
(dotted line) and the total number
of links (divided by $2000$; solid line) versus 
computer time. Here $N=1000, J=0.005$ and $\beta=0.020$.}
\end{figure}
What is even more interesting, one observes a 
strong (anti)correlation between the
wealth inverse participation ratio and the total number of links (see 
Fig.~\ref{fig:Y2vsL}). The periods of relatively low participation ratio and
large number of inter-node connections alternate with 
periods where participation
ratio is large and the number of links small. Increasing 
the number of trading links
apparently reduces ``social disparities''. Of course, 
this remark should not be taken too
seriously, the frequency of the regime changes is 
too rapid to be an image of the
behavior of actual markets. However, the trend is of interest.

\section{Discussion and conclusion}
\label{sect_CONCLUSIONS}
We have introduced a class of models in which agents perform trades
and influence the associated network of interactions.
We find that these adaptive network systems spontaneously go to a unique steady state,
and that several very distinct behaviors arise depending on
the parameters defining the models.
When no lower cut-off is imposed on agent wealth, the poor go
into a spiral of poverty and disconnect from the
network which ``collapses''; furthermore this is a cascading process
so that rapidly nearly all individuals reach this situation.
When instead a minimum wealth is enforced, 
the overall system reaches
a critical state where wealth and connectivity distributions
have power-law tails; this critical behavior is generic, no 
fine tuning of parameters is necessary. In this
critical steady state, the heterogeneity or ``differences'' in 
agent wealth depends on the trade intensity, parametrized
in our model by a coupling $J/\sigma^2$. For large $J/\sigma^2$, the wealth
circulates rapidly, and differences in wealth
are small. On the contrary when $J/\sigma^2$ is small, wealth differences
are large, and in fact for $J/\sigma^2$ small enough, one goes into a ``condensed''
phase where a finite fraction of the wealth is
held by just a few agents. 
Interestingly, we find this phase
transition point to be the same as when the
network is quenched according to any law for the
degree distribution. Not surprisingly, we have also found that the
wealth and the network dynamics lead to large correlated
fluctuations; in particular, the total wealth tends to 
be lowest when the network is the densest. 

The occurence of power laws in wealth distributions, usually referred
to as Pareto's law~\cite{pareto89,mandelbrot60,souma01}, has 
been empirically observed in many
economic contexts. Since such systems almost always involve
adaptive networks, it would be of major interest 
to extend those observations to the properties of the
underlying networks. Our model suggests not only that
these networks will be characterized by power laws, 
but that the wealth and network properties will be 
strongly correlated. In situations where regulation of such
behavior is considered necessary, policies may focus 
on the network ``rules'' rather than attempting
to regulate wealth directly; these policies might involve
introducing fees or subsidies for different kinds of
trades. Clearly in realistic situations, there may be
other features to take into account such as geographic
influences on the adaptive network dynamics. One may have to 
also consider social trends such as spontaneous assortativity
formation in trading networks. It seems to us in particular
that sufficient assortativity may prevent the spiral
of poverty formation when no minimum wealth is imposed.
More generally, many of these issues extend far beyond
economic adaptive networks: food-webs, 
transportation networks, or social networks all lead
to similar questions.

{\bf Acknowledgments---}
This work was supported
by the EEC's FP6 Marie Curie RTN under
contract MRTN-CT-2004-005616 (ENRAGE: European
Network on Random Geometry), 
by the EEC's IST project GENNETEC - 034952, by the
Marie Curie Actions Transfer of Knowledge project ``COCOS'',
Grant No. MTKD-CT-2004-517186, and by the Polish Ministry of Science
and Information Society Technologies Grant 1P03B-04029 (2005-2008).
The LPT and LPTMS are Unit\'e de Recherche de
l'Universit\'e Paris-Sud associ\'ees au CNRS.


\begin{thebibliography}{99}
\bibitem{gross} T. Gross and B. Blasius, arXiv:0709.1858.
\bibitem{zimmermannEguiluz04} M.G. Zimmermann, V.M. Eguiluz and 
M. San Miguel, Phys. Rev. E 69, 065102 (2004).
\bibitem{holmeNewman06} P. Holme and M.E.J. Newman, 
Phys. Rev. E 74, 056108 (2006).
\bibitem{kozmaBarrat08} B. Kozma and A. Barrat, Phys. Rev. E 77, 016102 (2008).
\bibitem{grossLima06} T. Gross, C.D. D'Lima and B. Blasius, 
Phys. Rev. Lett 96, 208701 (2006).
\bibitem{barabasiAlbertxx} L. Barabasi and R. Albert, Science 286, 509 (1999).
\bibitem{parkNewmanxx} J. Park and M.E.J. Newman, 
Phys. Rev. E 68, 026112 (2003).
\bibitem{pareto89} V. Pareto, Reprinted as a volume of \emph{Oeuvres
Compl\`etes} (Droz, Geneva), 1896 (1989).
\bibitem{mandelbrot60} B.B. Mandelbrot, 
Int. Eco. Rev., vol 1, 79 (1960).
\bibitem{souma01} W. Souma, Fractals, vol. 9, No. 3, 463 (2001).
\bibitem{bouchaudMezardXX} J.-P. Bouchaud and M. M\'ezard, 
Physica A: Statistical Mechanics and its Applications, vol. 282, 536 (2000).
\bibitem{garla} D. Garlaschelli et al., 
Eur. Phys. J. B 57, 159 (2007); for an earlier
paper by the same group devoted to adaptive networks see 
G. Caldarelli, A. Capocci and D. Garlaschelli, Nature Physics 3, 
813 (2007).
\bibitem{bouchaudMezardXX2} J.-P. Bouchaud and M. M\'ezard, 
J. Phys. A: Math. Gen. 30, 7997 (1997).

\end{thebibliography}

\end{document}